\documentclass[prl,a4paper,superscriptaddress,twocolumn,showpacs,amsmath,amssymb,floatfix]{revtex4-1}
\usepackage{amsfonts}
\usepackage{graphicx}

\begin{document}
\title{Multifractality of eigenfunctions in spin chains}
\author{Y.Y Atas and E. Bogomolny}
\affiliation{Univ. Paris-Sud, CNRS, LPTMS, UMR8626,  91405 Orsay, France}
\pacs{75.10.Jm, 75.10.Pq, 05.50.+q}
\date{\today}

\begin{abstract}
We investigate different one-dimensional quantum spin-$\tfrac{1}{2}$ chain models  and by combining analytical and numerical calculations prove  that their ground state wave functions in the natural spin basis are multifractals with, in general, non-trivial fractal dimensions.  
\end{abstract} 

\maketitle


-- \textit{One-dimensional quantum spin chains} are among the oldest and the most investigated fundamental models in physics. Introduced as toy-models of magnetism \cite{heisenberg}, they quickly became a paradigm of quantum integrable models (see e.g.  \cite{bethe}-\cite{baxter}). A prototypical  example  is the XYZ Heisenberg model \cite{heisenberg}  for $N$ spins-$\tfrac{1}{2}$  in external fields with periodic boundary conditions  
\begin{eqnarray}
\mathcal{H}=-\sum_{n=1}^N\Big [\frac{1+\gamma}{2} \sigma_n^{x}\sigma_{n+1}^x&+&\frac{1-\gamma}{2} \sigma_n^{y}\sigma_{n+1}^y 
+\frac{\Delta}{2}  \sigma_n^{z}\sigma_{n+1}^z\nonumber\\ &+&\lambda \sigma_n^z+\alpha \sigma_n^{x}\Big ]
\label{general_H}
\end{eqnarray} 
and its various specifications for different values of parameters. $\sigma_n^{x,y,z}$  are the Pauli matrices at site $n$. 

In the natural basis of $z$-components of each spin, $|\vec{\sigma}\rangle = |\sigma_1,\ldots,\sigma_N\rangle $ where $\sigma_j=\pm 1$, any Hamiltonian of $N$ spins-$\tfrac{1}{2}$ is represented by a $M\times M$ matrix with  dimension $M=2^N$.   Many different methods were developed  to  determine  exact spectra of such matrices \cite{bethe}-\cite{baxter}. 
The calculation of eigenfunctions is more involved. A wave function of $N$ spins-$\frac{1}{2}$ in the  spin-$z$ basis can be written as 
\begin{equation}
\mathbf{\Psi}=\sum_{\vec{\sigma}} \Psi_{\vec{\sigma}} | \vec{\sigma} \rangle
\label{psi_expansion}
\end{equation}
where the summation is taken over all $M=2^N$ configurations with $\sigma_j=\pm 1$. In general, coefficients $\Psi_{\vec{\sigma}}$ can be found only after the matrix diagonalization which for large $N$ is a hard numerical problem. Even in integrable cases  eigenfunctions of spin chains look erratic (cf. figures  below) and their structure is not well understood. 

The purpose of this letter is to prove that ground state (GS) wave functions for different one-dimensional spin chain models are multifractals in the spin-$z$ basis. 

-- \textit{Multifractality} is a general notion introduced to characterize strong and irregular fluctuations of  various quantities  \cite{mandelbrot}-\cite{stanley}.  For eigenfunctions like in \eqref{psi_expansion} one uses the following definition   (see e.g. \cite{mirlin} and references therein). Let $S_R(q,M)$ be the R\'enyi entropy for an eigenfunction \eqref{psi_expansion} of a matrix of finite  size $M$
\begin{equation}
S_R(q,M)=-\frac{1}{q-1}\ln \Big (\sum_{\vec{\sigma}}|\Psi_{\vec{\sigma}} |^{2q}\Big )
\label{renyi_entropy}
\end{equation}  
with   normalized coefficients  $\Psi_{\vec{\sigma}}$, $\sum_{\vec{\sigma}}|\Psi_{\vec{\sigma}} |^{2}=1$. 

Fractal dimensions, $D_q$, are defined  from the behaviour of the R\'enyi entropy \eqref{renyi_entropy} in the limit $M\to\infty$ \cite{mirlin}
\begin{equation}
D_q=\lim_{M\to \infty} \frac{S_R(q,M)}{\ln M}\ .
\label{fractal}
\end{equation}
The case when $D_q$ is a non-linear function of $q$ corresponds to a multifractal irregular behaviour. 

Fractal dimensions give a concise description of wave function moments and are important characteristics  of  eigenfunctions but it seems that they  were overlooked in previous studies. For example, the R\'enyi  and Shannon entropies are calculated for GS wave functions of certain spin chains in \cite{pasquier_1}--\cite{latorre} but  terms linear in $\ln M$ which determine $D_q$ in \eqref{fractal} were regularly ignored and only  next-to-the leading terms  have been investigated as it is usual in  conformal field theories. To the best of authors' knowledge  only a recent paper \cite{sierra} briefly mentioned the existence of  multifractality in a spin chain model. 

The simplest method to find fractal dimensions is the direct numerical calculation of the GS wave function for different number of spins and a subsequent extrapolation of the R\'enyi entropy for large $M$.  Definition \eqref{fractal} is well suited for positive $q$. For many problems (but not for all) fractal dimensions  can   be calculated also for negative $q$  \cite{mandelbrot_2}, \cite{riedi}. Of course, if certain coefficients in \eqref{psi_expansion} are zero due to an exact  symmetry, they are not included in  the calculation of the R\'enyi entropy \eqref{renyi_entropy} for $q\leq 0$. 

In what follows we investigate various specifications of the Heisenberg model \eqref{general_H} and by combining  numerical and analytical methods demonstrate that the multifractality of GS is a generic property of all of them. We choose $\gamma\geq 0$ and $\alpha\geq 0$ to ensure  off-diagonal terms of Hamiltonian matrices \eqref{general_H}  to be non-positive which,  by the Perron-Frobenius theorem, implies that  coefficients $\Psi_{\vec{\sigma}}$  in \eqref{psi_expansion} for the GS wave function are non-negative. Other  parameters are  such that  GS wave functions of  the XY models (with $\Delta=0$) are ferromagnetic and for the XYZ models they are anti-ferromagnetic.


-- \textit{The quantum Ising model in transverse field} \cite{pfeuty} is a standard model of quantum phase transitions \cite{sachdev}. It corresponds to the Hamiltonian \eqref{general_H} with $\Delta=\alpha=0$ and $\gamma=1$. Its spectrum can be found analytically  by the Jordan-Wigner transformation \cite{lieb} and coefficients $|\Psi_{\vec{\sigma}}|^2$  are given by the determinant of $N\times N$ matrices \cite{lieb}, \cite{pasquier_2}. 

\begin{figure}[t]
\begin{center}
\includegraphics[ width=.9\linewidth,clip]{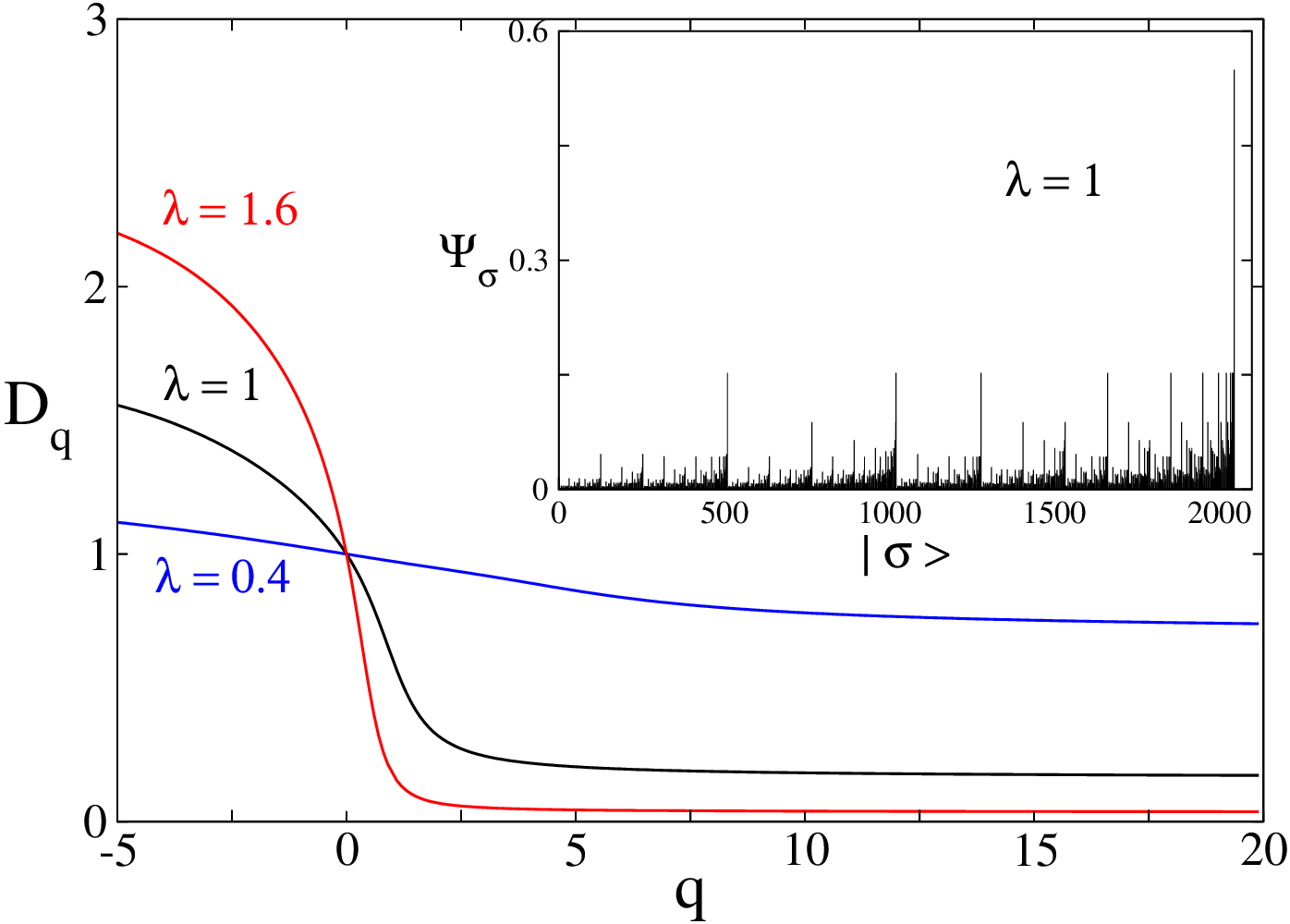} 
\end{center}
\caption{(color online). Fractal dimensions of GS for the quantum Ising model. Red line: $\lambda=1.6$,  black line: $\lambda=1$, blue line: $\lambda=0.4$.  Inset: GS  coefficients for $\lambda=1$ and $N=11$. The abscissa axis here and in other figures is integer binary code for $\vec{\sigma}$, $x=\sum_{n=1}^{N} 2^{n-2}(1+\sigma_n)$.}
\label{fig_Ising}
\end{figure}

Fractal dimensions of GS wave function for this model  computed numerically from linear extrapolation of the R\'enyi entropy  with $N=3-11$ are presented in Fig.~\ref{fig_Ising} for a few values of transverse field $\lambda$. The curves of $D_q$ as a function of $q$ for all models   have the same characteristic form as for other fractal measures \cite{kadanoff}.  In particular,  when $q\to \pm \infty$  they tend to  well defined limits $D_{\pm \infty}$.          

\begin{figure}
\begin{center}
\includegraphics[ width=.7\linewidth,clip]{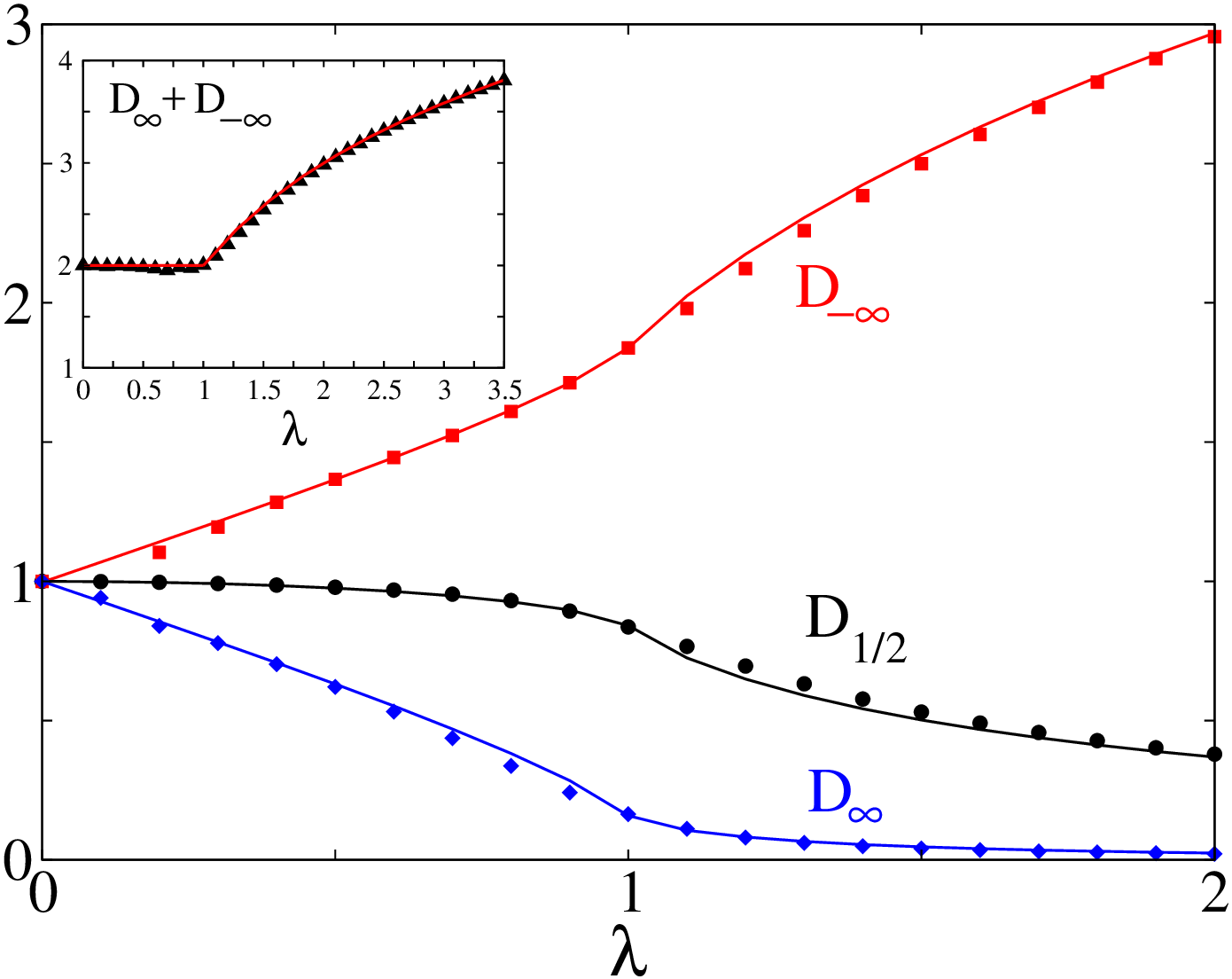} 
\end{center}
\caption{(color online). Exact fractal dimensions \eqref{D_pm} for the quantum Ising model: $D_{\infty}$ (blue line), $D_{1/2}$ (black line), and $D_{-\infty}$ (red line).  Results of numerical calculations are indicated by symbols of the same color. Inset: $D_{\infty}+D_{-\infty}$ calculated numerically (black triangles)  in comparison with  Eq.~\eqref{D_minus} (red solid line). }
\label{dimension_Ising}
\end{figure}

Generalizing  results of \cite{pasquier_2}, one gets the exact expressions for limiting values $D_{\pm \infty}$ and for $D_{1/2}$ 
\begin{eqnarray}
&&D_{\pm \infty}(\lambda)=\frac{1}{2}-\frac{1}{2\pi \ln 2}\int_0^{\pi}\ln \Big [1\pm \frac{\lambda-\cos u}{\sqrt{R(\lambda,u)}} \Big ]\mathrm{d}u , \label{D_pm}\\
&&D_{1/2}(\lambda)=1-D_{\infty}\Big (\frac{1}{\lambda}\Big ) ,\quad R(\lambda,u)=1-2\lambda \cos u +\lambda^2. \nonumber 
\end{eqnarray}
These formulas prove that fractal dimensions of quantum Ising model are non-trivial. In Fig.~\ref{dimension_Ising} these exact expressions are plotted together with numerically calculated points for different $\lambda$. $D_{\pm \infty}$ are obtained by a fit $D_{\pm \infty}+a/q+b/q^{2}$  fo large $q$ parts of curves similar to Fig.~\ref{fig_Ising}.  The good agreement between Eqs.~\eqref{D_pm} and numerics  shows that though we obtain fractal dimensions from relatively small number of spins our results are reliable.  

The above curves are qualitatively the same for non-critical and critical (that is $\lambda=1$) quantum Ising model. Nevertheless, as illustrated in the inset of Fig.~\ref{dimension_Ising} the sum of $D_{\infty}$ and $D_{-\infty}$ has clear singularity in the critical point in accordance with the relation 
\begin{equation}
D_{-\infty}(\lambda)+D_{\infty}(\lambda)=\left \{ \begin{array}{cc}2, &|\lambda|<1\\2+\frac{\ln |\lambda|}{\ln 2}, &|\lambda|>1\end{array}\right .
\label{D_minus}
\end{equation}
which follows from \eqref{D_pm}. It means that criticality can be observed in fractal dimensions of GS.     


-- \textit{The XY model} is a specification of \eqref{general_H} with $\Delta=\alpha=0$ and $\gamma\neq 1$.  Similar to the quantum Ising model this model is also integrable by the Jordan-Wigner transformation \cite{lieb}, \cite{pasquier_2} but the structure of its GS is more complicated. An interesting special case  is $\lambda=\lambda_f$ where $\lambda_f=\sqrt{1-\gamma^2}$.   
It is known \cite{kurmann} that at that field the XY model has two exact factorized GS wave functions
\begin{equation}
\Psi=\prod_{n=1}^N(\cos \theta |1\rangle_n \pm \sin \theta |-1\rangle_n ),\quad \cos^2 2\theta=\frac{1-\gamma}{1+\gamma}.
\label{factorized}
\end{equation}
\begin{figure}[b]
\begin{center}
\includegraphics[width=.9\linewidth,clip]{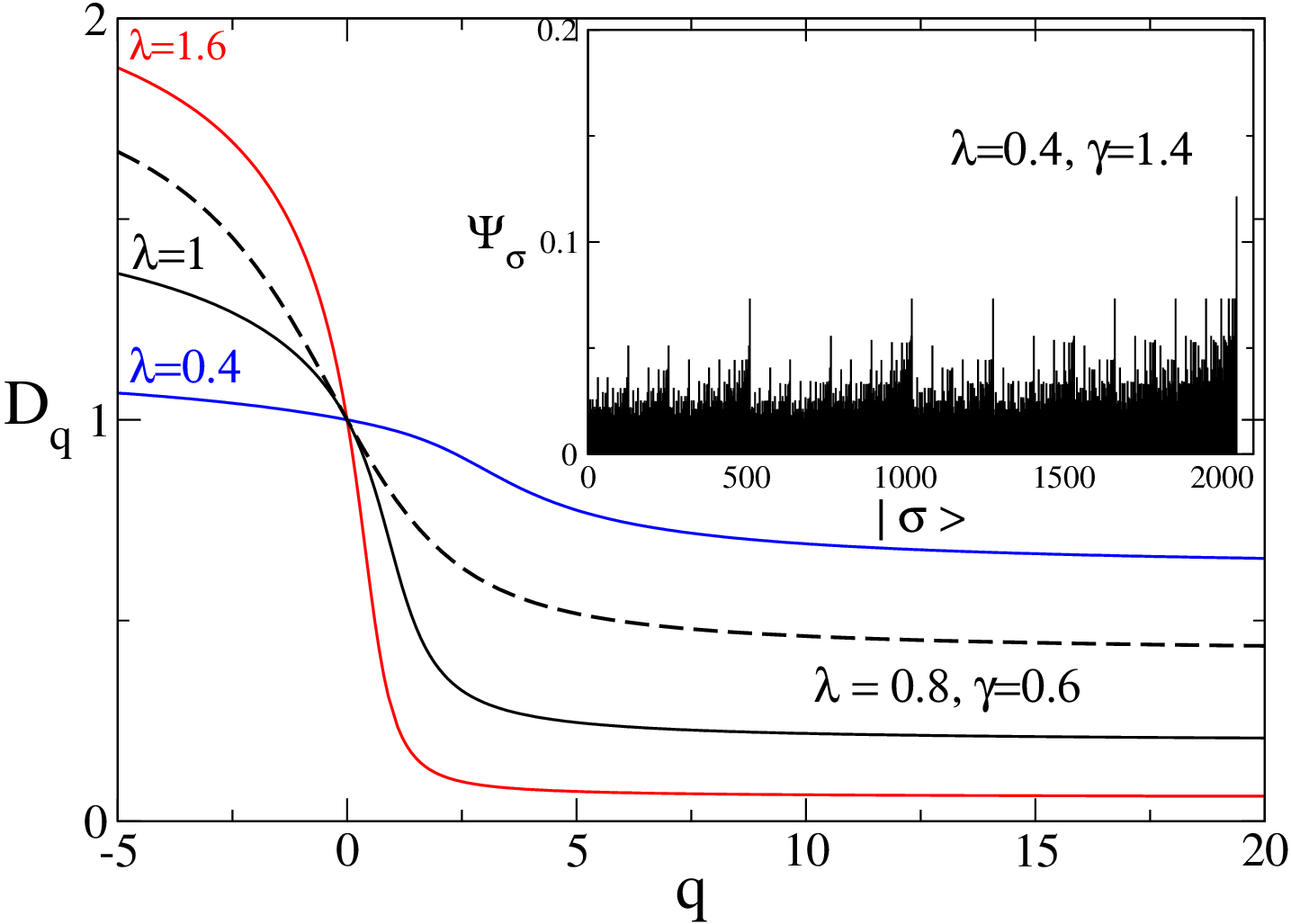} 
\end{center}
\caption{(color online). Fractal dimensions of GS for the XY model with anisotropy $\gamma=1.4$.  Red line:  $\lambda=1.6$, black line:  $\lambda=1$, blue line:  $\lambda=0.4$. Dashed black line shows for comparison  the exact fractal dimensions \eqref{xy_exact}  for $\lambda=0.8$ and $\gamma=0.6$. Inset: GS  coefficients for $\lambda=0.4$, $\gamma=1.4$  and $N=11$. }
\label{fig_XY}
\end{figure}
For states with definite parity and $\lambda=\lambda_f$ fractal dimensions  are described by a formula
\begin{equation}
D_q=-\frac{\ln( \cos^{2q}\theta+\sin^{2q}\theta)}{(q-1)\ln 2}
\label{xy_exact}
\end{equation}
 indicated for $\gamma =0.6$ and $\lambda=0.8$ by dashed black line in Fig.~\ref{fig_XY}. This example proves that at least at the factorizing field fractal dimensions of the GS wave function do exist and correspond to the well-investigated case of binomial measures \cite{mandelbrot}. 

As for $\lambda^2+\gamma^2<1$ there exist many crossings of lowest states with different parity,  for numerical calculations (performed as in the Ising model) we choose  $\lambda$ and $\gamma$ outside the unit circle, $\lambda^2+\gamma^2>1$.  The results are presented in Fig.~\ref{fig_XY} and are qualitatively similar to the Ising model.

One may argue that the limiting values, $D_{\infty}$ and $D_{-\infty}$ as in the quantum Ising model  should correspond to configurations with, respectively, all spins up and all spins down, and, consequently, are expressed  similar to \eqref{D_pm} as
\begin{equation}
D_{\pm }(\lambda, \gamma)=\frac{1}{2}-\frac{1}{2\pi \ln 2}\int_0^{\pi}\ln \Big [1\pm \frac{\lambda-\cos u}{\sqrt{R_{-}(\lambda, \gamma,u)}} \Big ]\mathrm{d}u 
\label{D_limit_XY}
\end{equation}
with $R_{-}(\lambda, \gamma,u)=(\lambda- \cos u)^2 +\gamma^2 \sin^2 u$. But for small $\lambda$ the minimal contribution is instead given  by the anti-ferromagnetic N\'eel configuration with alternating spins, $\sigma_n=(-1)^n$. Using the asymptotics of the block Toeplitz matrices \cite{widom} we get
\begin{eqnarray}
&&D_{\mathrm{Neel}}(\lambda, \gamma)=\label{D_Neel_XY}\\
&& \frac{3}{4}-\frac{1}{2\pi \ln 2}\int_0^{\pi/2}\ln \Big [1- \frac{\lambda^2+\gamma^2-(1+\gamma^2)\cos^2 u}{\sqrt{R_{+}(\lambda, \gamma,u)R_{-}(\lambda, \gamma,u)}}\Big ]\mathrm{d}u 
\nonumber
\end{eqnarray}
where $R_{+}(\lambda, \gamma,u)=(\lambda+ \cos u)^2 +\gamma^2 \sin^2 u$.
\begin{figure}[t]
\begin{center}
\includegraphics[width=.7\linewidth,clip]{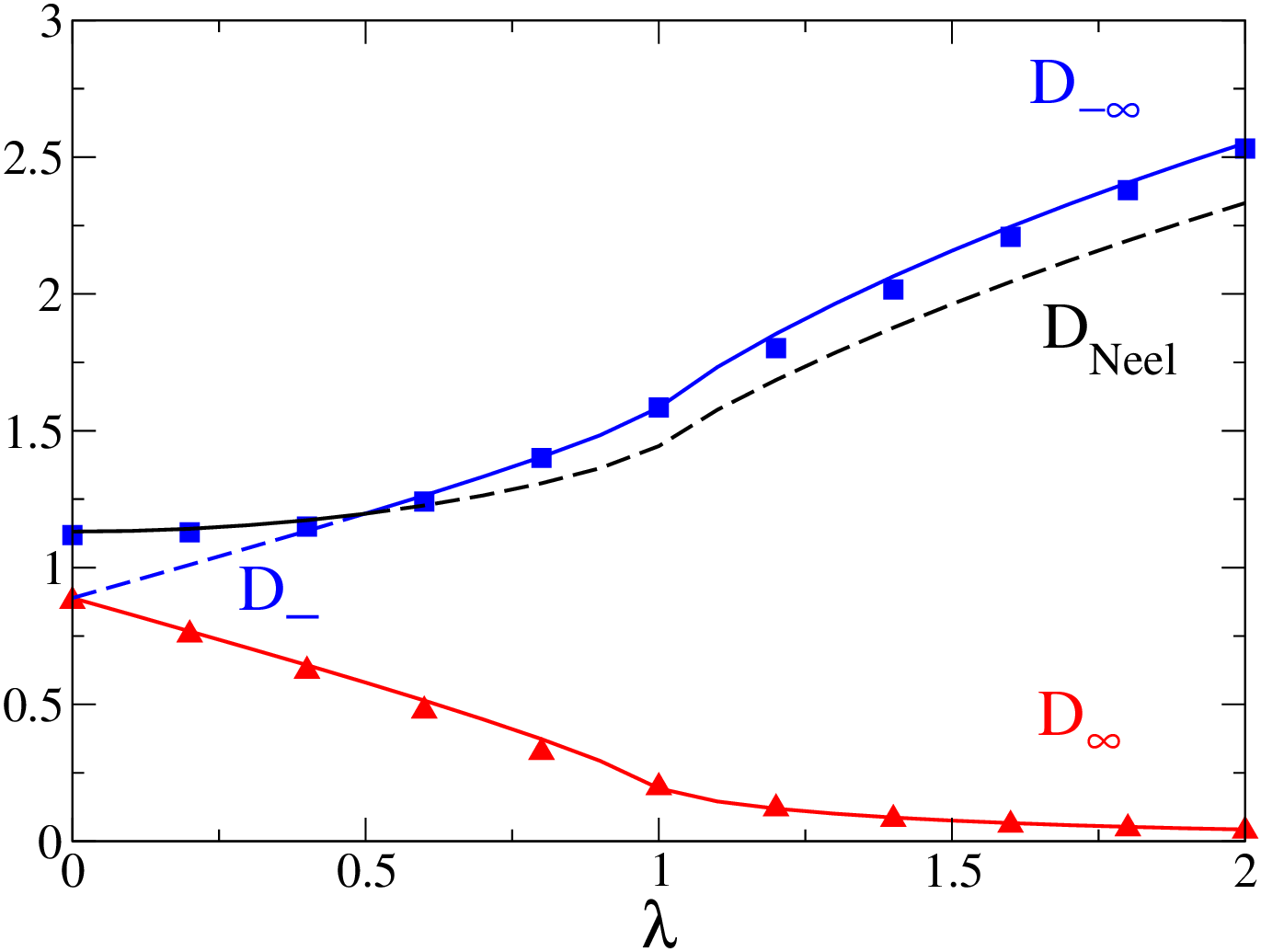} 
\end{center}
\caption{(color online). Asymptotic fractal dimensions for the XY model with anisotropy $\gamma=1.4$ versus  transverse field. Red and blue lines indicate $D_{\pm}$ in \eqref{D_limit_XY}. Black line is the contribution of the N\'eel configuration \eqref{D_Neel_XY}. When contributions become sub-dominant they are indicated by dashed lines of the same color. Blue squares and red triangles are respectively $D_{-\infty}$ and $D_{\infty}$ calculated  as for the Ising model. }
\label{dimension_XY}
\end{figure}
This result for $\gamma=1.4$ is presented in Fig.~\ref{dimension_XY} by the black line. 
When $D_{\mathrm{Neel}}>D_{-}$, $D_{-\infty}=D_{\mathrm{Neel}}$, otherwise  $D_{-\infty}=D_{-}$. For $\gamma=1.4$ these curves intersect at $\lambda\approx 0.4982$ and $D_{-\infty}$ has the form indicated in Fig.~\ref{dimension_XY} by solid  blue and black lines. Numerical results agree well with this prediction. 

\begin{figure}[t]  
\begin{center}
\includegraphics[width=.9\linewidth,clip]{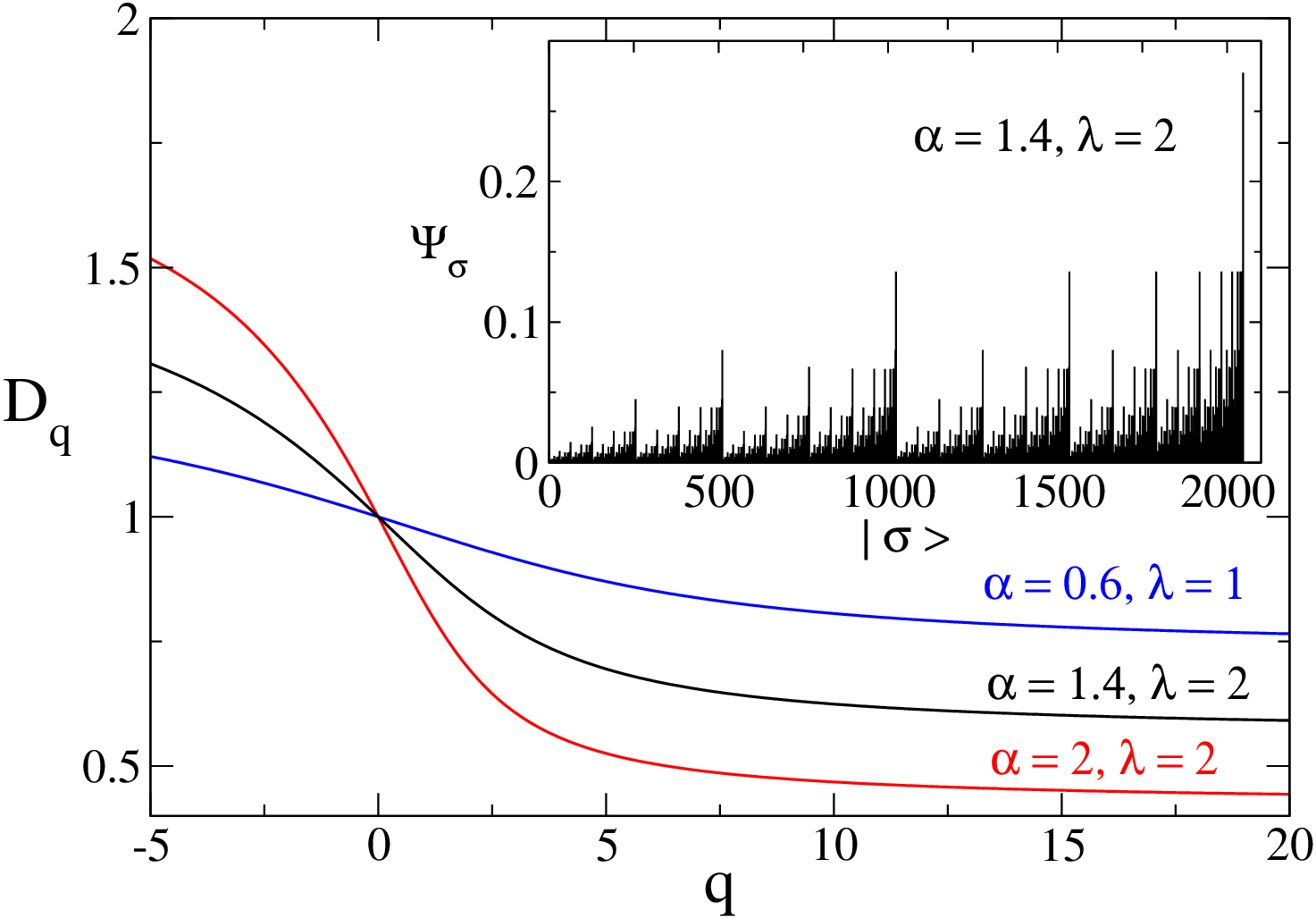} 
\end{center}
\caption{(color online). Fractal dimensions for the quantum Ising model in transverse ($\lambda$) and longitudinal ($\alpha$) fields.  Red line:  $\alpha=2, \; \lambda=2$, black line:  $\alpha=1.4,\;\lambda=2$, blue line:  $\alpha=0.6,\;\lambda=0.4$. Inset: GS  coefficients for $\alpha=1.4,\;\lambda=2$,  and $N=11$.}
\label{fig_Ising_fields}
\end{figure}

-- \textit{The Ising model in transverse and longitudinal fields} is obtained by adding to  the quantum Ising model a longitudinal field $\alpha$. 
For non-zero $\alpha$ the only known integrable case corresponds to $\lambda=1$ \cite{zamolodchikov}. This  model attracts recently   wide attention as certain consequences of its integrability have been checked experimentally in the cobalt niobate ferromagnet  \cite{experiment}. In Fig.~\ref{fig_Ising_fields} the fractal dimensions for a few values of both  fields are presented. 

-- \textit{The XXZ model} in zero fields  is  a particular case of the Heisenberg model \eqref{general_H} with $\gamma=\lambda=\alpha=0$ and $\Delta\neq 0$. Due to the conservation of the $z$ component of the total spin, $S_z=\sum_n\sigma_n^{z}$, its Hamiltonian   can be diagonalized  in subspace with fixed $S_z$.   The model is soluble by the coordinate Bethe anzatz \cite{bethe}, \cite{yang}, \cite{orbach} and has a rich  phase  diagram (see e.g.  \cite{mattis}). 

As a reference  we use  $\Delta=-\tfrac{1}{2}$ called combinatorial point.  From the Ra\-zu\-mov--Stroganov conjecture \cite{razumov} proved in \cite{cantini} it follows that at such $\Delta$ and odd $N=2R+1$ the following statements are valid: (i) the GS energy is $-3N/4$, (ii) the largest coefficient in the expansion \eqref{psi_expansion} (the one for the N\'eel configuration)  equals 
\begin{equation}
\Psi_{\mathrm{max}}^{-1}=\frac{3^{R/2}}{2^R} \frac{2\cdot 5\ldots (3R-1)}{1\cdot 3\ldots (2R-1)}\ ,
\end{equation}
(iii) the smallest coefficient corresponding to a half consecutive spins up and other spins down is
$\Psi_{\mathrm{min}}^{-1}=\Psi_{\mathrm{max}}^{-1}A_R$,
and (iv) $\sum_{\vec{\sigma}}\Psi_{\vec{\sigma}}=3^{R/2}$. 
Here $A_R$ is the number of alternating sign matrices  \cite{asm}.

\begin{figure}[t]   
\begin{center}
\includegraphics[width=.9\linewidth,clip]{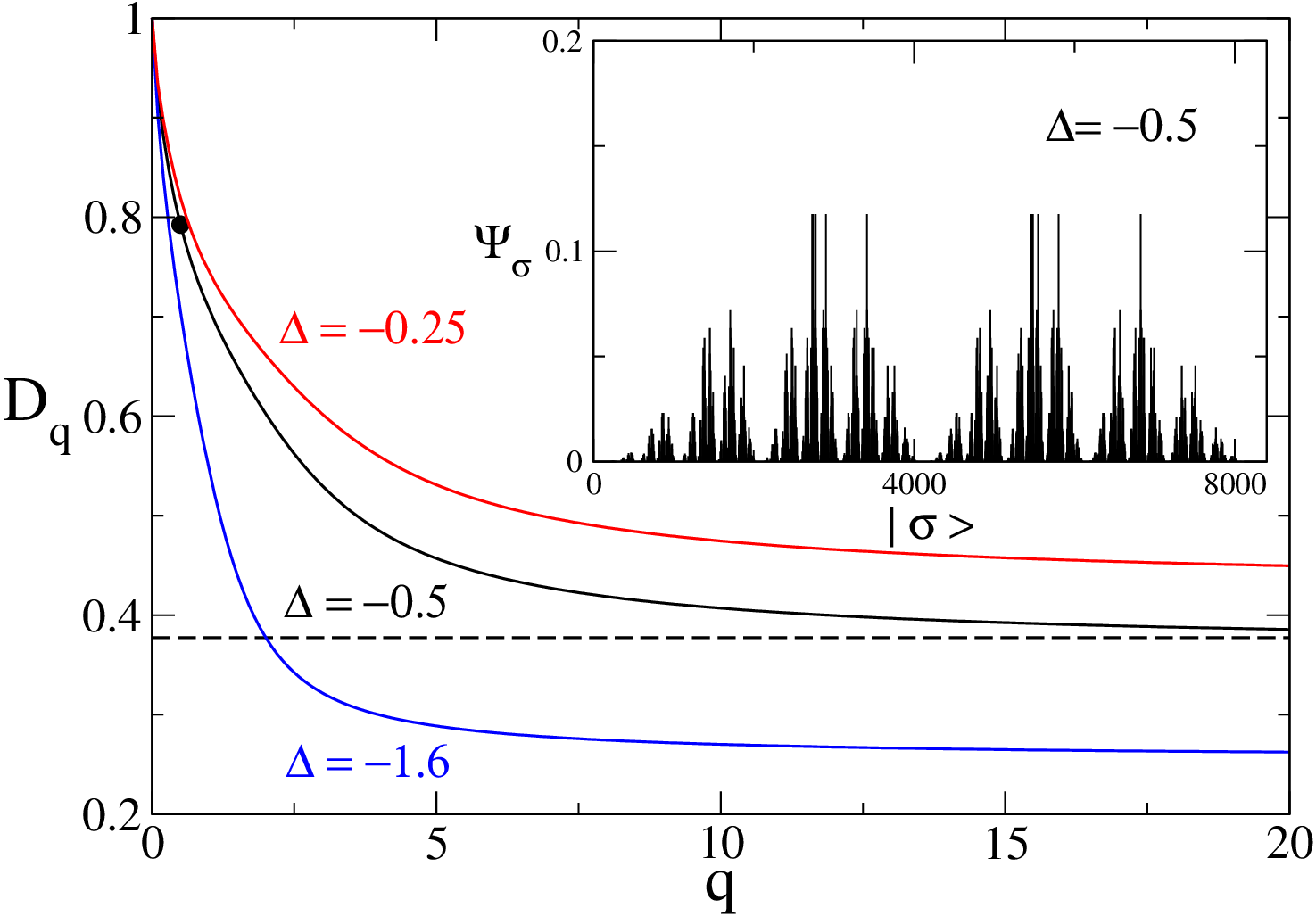} 
\end{center}
\caption{(color online). Fractal dimensions for the XXZ model in zero fields. Red line:  $\Delta=-0.25$, black line:  combinatorial point $\Delta=-0.5$, blue line:  $\Delta=-1.6$. Inset: GS  coefficients for $\Delta=-0.5$ and $N=13$. Dashed line indicates theoretical prediction for $D_{\infty}$ \eqref{razumov_points} at  $\Delta=-0.5$. Black circle is the value of $D_{1/2}$ \eqref{razumov_points} at this point. }
\label{fig_XXZ}
\end{figure}

These formulas prove that for $\Delta=-\tfrac{1}{2}$ fractal dimensions $D_{\infty}$ and $D_{1/2}$ are explicitly known
\begin{equation}
D_{\infty}=\frac{3\ln 3}{2\ln 2}-2 \approx 0.377,\quad D_{1/2}=\frac{\ln 3}{2\ln 2}\approx 0.792\, .
\label{razumov_points}
\end{equation} 
When $R\to\infty$, $\ln A_R= R^2\ln (3\sqrt{3}/4)  +\mathcal{O}(R)$.  Such quadratic in $N$  behaviour is a particular case of the emptiness formation probability of a string of $n$ aligned spins with $n\sim N$ \cite{emptiness}. This asymptotics  means that negative moments of the GS wave function  in anti-ferromagnetic case  require a scaling different from \eqref{fractal} and hence will not be considered here.  The fractal dimensions for a few values of parameter $\Delta$  are presented in Fig.~\ref{fig_XXZ}. The numerical calculations were performed by extrapolation of
the R\'enyi entropy separatively for odd and even $N=3-13$. With available precision, fractal dimensions for odd and even $N$ are the same but sub-leading terms in the R\'enyi entropy  \eqref{renyi_entropy} are different.  


-- \textit{The XYZ model}   differs from the XXZ model by anisotropy $\gamma\neq 0$ in the $(x,y)$ plane. In zero fields its GS wave function  has been found in \cite{baxter}.  A soluble example  with factorized GS  \eqref{factorized} is $\alpha=0$, $\lambda=\lambda_f$  \cite{kurmann} where
\begin{equation}
\lambda_f=\sqrt{(1-\Delta)^2-\gamma^2}, \quad  \cos^2 2\theta=\frac{1-\gamma-\Delta}{1+\gamma-\Delta}\ .
\end{equation}
At this field GS wave function  corresponds to the binomial measure  and fractal dimensions are given by \eqref{xy_exact}. 

\begin{figure}[t]
\begin{center}
\includegraphics[width=.9\linewidth,clip]{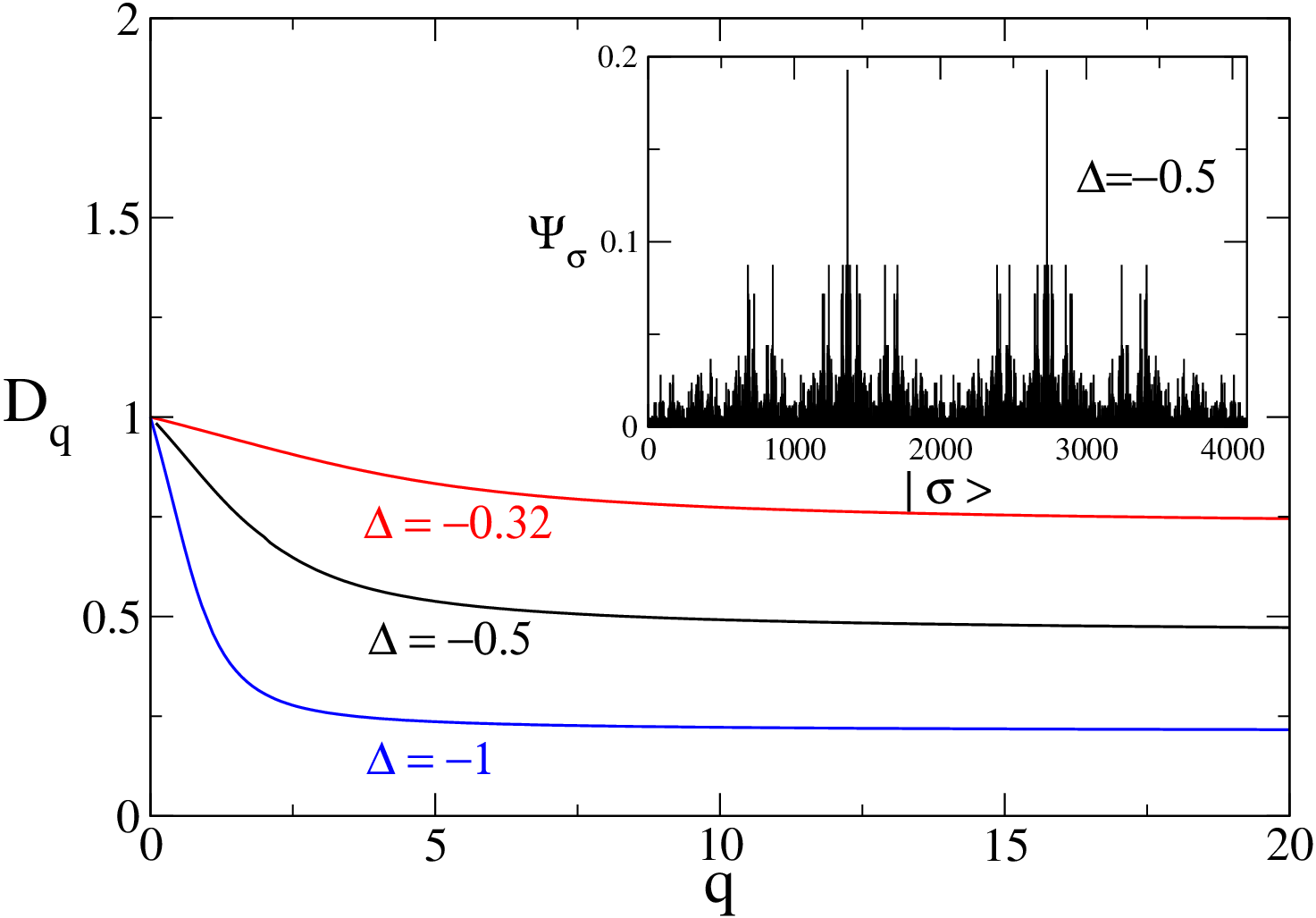} 
\end{center}
\caption{(color online). Fractal dimensions for the XYZ model with anisotropy $\gamma=0.6$.  Red line: combinatorial point $\Delta=-0.32$, black line:  $\Delta=-0.5$, blue line:  $\Delta=-1$. Inset: GS coefficients for $\Delta=-0.5$, $\gamma=0.6$,  and $N=12$. }
\label{fig_XYZ}
\end{figure}

The combinatorial point for the XYZ model  in zero fields where an additional information about GS  is known (or conjectured) is  $\Delta=(\gamma^2-1)/2$ \cite{rasumov_2}. In Fig.~\ref{fig_XYZ} we present fractal dimension in the zero fields XYZ model with $\gamma=0.6$ for a few values of $\Delta$ including the combinatorial point. Qualitatively the curves are similar to the XXZ model and to $D_q$ with positive $q$ for the XY models. 

-- \textit{Summary}. Wave functions are the most fundamental objects of any quantum-mechanical model. For  many-body problems their structure is complicated and  numerous  questions remain open. We consider  here practically all standard one-dimensional spin-$\tfrac{1}{2}$ models and demonstrate that their GS wave functions in the natural spin-$z$ basis are multifractals with, in general, non-trivial fractal dimensions.  For special values of parameters and/or  certain dimensions we get exact analytical formulas which prove rigorously the existence of fractal dimensions. In other cases we rely on numerical calculations. The multifractality in spin chains is a very robust phenomenon. It exists for integrable and non-integrable models, for ferro and anti-ferro magnetic states, as well as for critical and non-critical systems.

 A common point of all these models is that their $M\times M$ Hamiltonian matrix in spin-$z$ basis is such that in each row and  column there exist only $K\sim \ln M$ non-zero matrix elements of the same order. As $M\gg K$ this specific form resembles a tree structure with branching number $K$. As $K\to\infty$ when $M\to\infty$ and  it is known \cite{localization}, \cite{altshuler} that on a tree with large branching number (with other parameters fixed) the Anderson localization is  unlikely,   states   are delocalized. On the other hand, the full ergodicity on a tree is, in general, also improbable \cite{altshuler}. The remaining possibility corresponds to delocalized but not ergodic states, i.e. to multifractality, which may explain its ubiquity.  The multifractality  is related finally not with explicit randomness but with internal complexity of models considered. Such arguments are not restricted to one dimension and/or spin chains but can be applied to various many-body problems with local interactions and we conjecture that multifractality (i.e. a non-trivial scaling of different quantities with the number of particles) is a generic property of a large class of many-body models. 


-- \textit{Acknowledgements}. The authors are greatly indebted to  O. Giraud  and G. Roux for  numerous useful comments and valuable help in performing numerical calculations. We thank  B.~Altschuler for  fruitful discussions, O. Giraud for careful reading of the manuscript, and  G. Roux  for pointing out  Ref.~\cite{sierra}. This work was supported by the CFM Foundation-JP Aguilar grant.
 

\end{document}